# CHOLESTERIC LIQUID CRYSTALS WITH TUNABLE DEFECT INSIDE

A.H. Gevorgyan[1], K. B. Oganesyan [2,3*], E.A. Ayryan[3], M. Hnatic [4,5,6], Yu.V. Rostovtsev [7]

[1] Department of Physics, Yerevan State University, Yerevan,  Armenia
[2] Yerevan Physics Institute, Alikhanyan Br.2, 036, Yerevan, Armenia
[3] LIT, Joint Institute for Nuclear Research, Dubna, Russia
[4] Faculty of Sciences, P. J. Safarik University, Park Angelinum 9, 041 54 Kosice, Slovakia
[5] Institute of Experimental Physics SAS, Watsonova 47, 040 01 Kosice, Slovakia ˇ
[6] BLTP, Joint Institute for Nuclear Research, Dubna, Russia
[7] University of North Texas, Denton, TX, USA

[*] bsk@yerphi.am

**Abstract.** Peculiarities of the defect modes of cholesteric liquid crystals (CLCs) with an isotropic/anisotropic defect inside are investigated. The influence of the defect layer thickness and its anisotropy of refraction, the influence of the system thickness and of the defect layer position in the system, as well as the influence of the dielectric borders on the defect modes is investigated. It is shown that it is possible to change reflection at the defect modes in wide intervals and change the defect mode wavelength, by tuning the defect location and its thickness. Such a system possesses transmission asymmetry. Also, the CLC system thickness and the refraction coefficient of the medium bordering the CLC layer on its both sides have essential influence on the reflection at the defect mode and on the reflection frequency.

## I. INTRODUCTION

In recent years, considerable interest has been attracted to the photonic crystals (PCs) [1-4], which are a special class of artificial and self organizing structures with periodic changes of spatial dielectric properties in the scale of optical order of wavelength. Such media are also called *media with photonic band-gap* (PBG), since there is a zone of frequency in their transmission spectra, where light undergoes diffraction reflection on their periodical structure. The interest in PCs is conditioned both by their interesting physical properties and wide practical applications. As these structures are designed artificially or in a self assembled manner, they can be prepared with given properties, which lead to many challenging problems of theoretical and applied character. The optical elements constructed on the basis of PCs results in intelligent, multifunctional tunable optics, which possess such favorable traits, such as their compactness, small losses, high reliability and compatibility with other devices. Cholesteric liquid crystals (CLCs) are the most representative among the one dimensional (1D) chiral PCs, because they can spontaneously self organize their periodic structure, and their PBG (that exists only for circularly polarized light with the same handedness of the CLC helix), and they can be easily tuned over wide frequency intervals.

Recently, CLCs have drawn great interest to them due to their possibility of low-threshold laser generation at the edges of their PBG. Dowling et al. predicted [5] lasing at the band edge of photonic band gaps materials based on the argument that light slows down near the band edge and so spontaneous emission would be enhanced. The Dowling mechanism of lasing only applies to infinite or very long systems. In [6], it was discovered that CLCs are photonic band gap materials. In the same work it was discovered the lasing in finite systems, such as CLCs, where the lasing occurs in specific modes, which have very different lifetimes depending on how close they are to the band edge and this gives the selectivity to lase in long lived modes. And it should be noted that it is not the light with a low velocity that is involved, but essentially standing waves associated with standing resonances. Vigorous investigations in this area have been going on up to now (see [7] and references cited therein).

Besides, recently the CLC having various types of defects have been considered from the point of view of generating additional resonance modes in them and of investigating the possibilities of low-threshold laser generation at these modes. It is to be noted that CLCs with a defect in the structure possess a number of peculiarities which the isotropic 1D PCs lack (see below, section III). Recently, CLCs with an anisotropic (isotropic) defect were considered [8-25] similar to [32-93].

In this paper, we investigate (by numerical simulations) some new peculiarities of the defect modes in the CLC with an isotropic/anisotropic defect and found out different features of such a system. To investigate the influence of the dielectric borders, we study two cases: a) the system is sandwiched between the two half-infinite isotropic spaces with refractive indices $n_s$ equal to the CLC average refractive index given by $\bar{n} = \sqrt{(n_o^2 + n_e^2)/2}$ ($n_o = \sqrt{\varepsilon_1}$ and $n_e = \sqrt{\varepsilon_2}$ are the ordinary and extraordinary refractive indices of the local anisotropic structure; in this case the dielectric borders influence is minimum) and $\alpha = \bar{n}/n_s = 1$; and b) the system is in the vacuum and $\alpha = \bar{n}$.

## II. THE METHOD OF ANALYSIS

The problem is solved by Ambartsumian's layer addition modified method [13,19] adjusted to solution of such problems. A *CLC layer* with a defect can be treated as a multi-layer system: *CLC(1)-Defect Layer (DL)-CLC(2)* (Fig. 1).

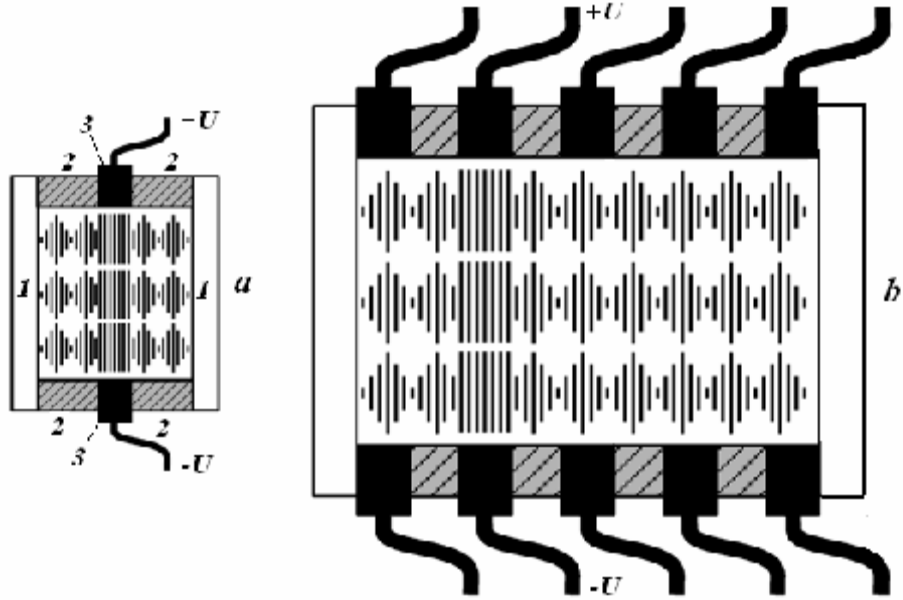

Fig. 1. A sketch diagram of a cell with a chiral liquid crystal with one defect: 1– glass substrates; 2—Teflon fillers; 3—metal electrodes.

According to Ambartsumian's layer addition modified method, if there is a system consisting of two adjacent (from left to right) layers, *A* and *B,* then the reflection transmission matrices of the system, $A+B$, viz. $\hat{R}_{A+B}$ and $\hat{T}_{A+B}$, are determined in terms of similar matrices of its component layers by the matrix equations:

$$\hat{R}_{A+B} = \hat{R}_A + \tilde{\hat{T}}_A \hat{R}_B \left[ \hat{I} - \tilde{\hat{R}}_A \hat{R}_B \right]^{-1} \hat{T}_A,$$
$$\hat{T}_{A+B} = \hat{T}_B \left[ \hat{I} - \tilde{\hat{R}}_A \hat{R}_B \right]^{-1} \hat{T}_A,$$
(1)

where the tilde denotes the corresponding reflection and transmission matrices for the reverse direction of light propagation, and $\hat{I}$ is the unit matrix. The exact reflection and transmission matrices for a finite *CLC layer* (at normal incidence) and a defect (isotropic or anisotropic) layer are well known [23, 24]. First, we attach the *DL* with the *CLC Layer (2)* from the left side, using the matrix Eqs (1). In the second stage, we attach the *CLC Layer (1)* with the obtained *DL-CLC Layer (2)* system.

## II. RESULTS AND DISCUSSION

We investigate reflection (transmission) spectra for either orthogonal linear or orthogonal circular polarizations. The ordinary and extraordinary refractive indices of the **DL** are taken to be $n_o^N = 1.522$ and $n_e^N = 1.74$, (these are the parameters of the nematic liquid crystal, E7, at $t = 25°C$ and at λ=0.59 mµ), and $n^d = 1.7$ (for the isotropic defect). The ordinary and extraordinary refractive indices of the **CLC layer** are taken to be $n_o = 1.4639$ and $n_e = 1.5133$, the **CLC layer** helix is right handed and its pitch is, $p = 0.42$ µm. These are the parameters of the **CLC cholesteryl-nonanoate–cholesteryl chloride–cholesteryl acetate** (20 : 15 : 6) composition, again at the temperature $T = 25^oC$. So, the light normally incident onto a single **CLC layer** – with the right circular polarization (RCP) – has a PBG, and the light with the left circular polarization (LCP) does not.

From the experimental point of view, fabricating a thin isotropic or anisotropic defect layer in the *CLC* structure is a hard task, though the problem theoretically is investigated in sufficient details. But it is not the case for the twist defect, which has been given substantially detailed account both theoretically and experimentally.

An interesting idea was proposed in Ref. [28]. Since liquid crystals are readily controlled, an anisotropic defect inside a *CLC* can be formed and its position can be changed, for example, with the aid of an external electrostatic field generated by a system of electrodes arranged along the *CLC* axis [28]. The external electrostatic field perpendicular to the helix axis produces aligning action on the *CLC* molecules (Fig. 1a). At a certain magnitude of the electric field ($E_{cr}$), the *CLC* molecules are aligned along the field lines, thus creating an anisotropic defect. The thickness of this defect layer is determined by the longitudinal size of the electrodes and by the applied voltage. Moreover, by applying voltage to different pairs of electrodes, it is possible to change the position of these defects in the system (Fig. 2b).

Below we investigate the effects of defect thicknesses and defect locations in the system on the reflection.

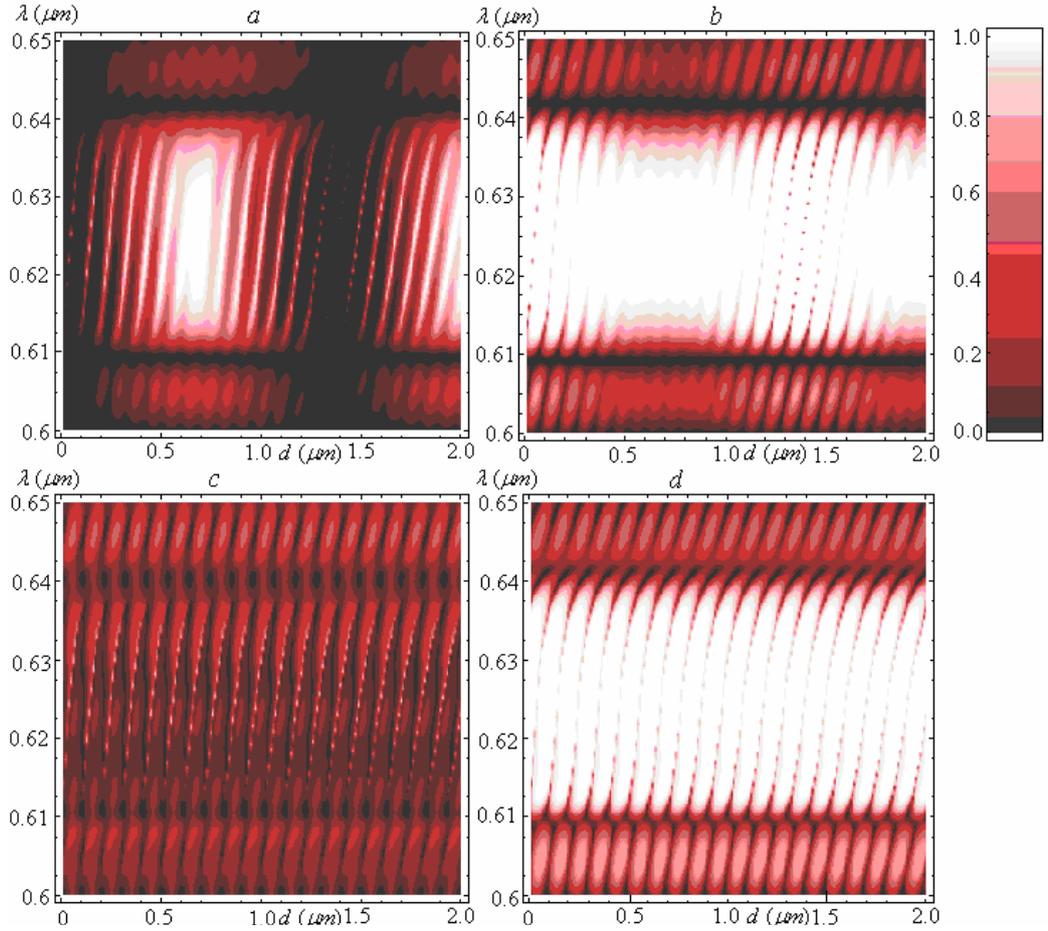

Fig. 2. The density plot of the reflection spectra as a function of defect layer thicknesses. *a* and *b* are for the anisotropic defects, and *c* and *d* are for the isotropic ones. The incident light is of the LCP (*a, c*) and RCP (*b, d*). $\alpha = 1$ (*a,b*) and $\alpha = \bar{n}$ (*c, d*). CLC layer thickness: $d=50p$.

In the Fig. 2, the density plot of the reflection spectra as a function of the defect layer thicknesses are presented. Figs. 2a, b corresponds to an anisotropic defect, and Figs. 2c, d correspond to an isotropic defect. Incident light is LCP (a, c), and RCP (b, d). The presence of a thin isotropic/anisotropic defect in the CLC structure creates a defect mode in the PBG. It manifests itself in the form of a hole in the reflection spectrum (in the PBG) for the light with the RCP, and in the form of a peak in the reflection spectrum for the light with the LCP. The defect mode has either donor or acceptor character depending on the optical thickness of the defect layer. The defect mode wavelength increases from a minimum to a maximum band gap value if the defect layer optical thickness increases, and at borders of the PBG two defect modes appear, then the longer-wavelength mode leaves the PBG and the shorter-wavelength mode moves into the PBG, as the defect layer thickness increases further. The reflection/transmission coefficient at the defect mode changes with oscillations if the defect layer thickness increases. In the case of the anisotropic defect the behavior is more complicated. For the anisotropic defect, if the layer thickness increases, the defect mode widths are also changed. The reflection of LCP light in the

PBG increases with oscillations, and when the defect layer thickness becomes of the order of 1.3 (i.e. the defect layer becomes a half-wave plate) the reflection of the two circular polarizations are nearly the same (except for weak difference is observed at the defect mode).

As mentioned above, an anisotropic defect in the CLC layer can be created and tuned by an external electrostatic field [25]. Consequently, one can change the location of the defect in the CLC by such field and, therefore, tune the position of the defect in the CLC system. Hence, it is necessary to study the influence of the defect layer location change in the CLC on the reflection spectra of the CLC.

In Fig. 3a we present the 3D plots of the reflectance coefficient, $R$, on the wavelength, $\lambda$, and on the reduced distance between the CLC right border and the defect layer (on $d/p$, where $d$ is the distance between the CLC right border and the DL); and in Fig. 3b the transmittance coefficient, $T$, on $\lambda$ and on $d/p$ are presented. The incident light is of the LCP (a) and RCP (b). The defect is an isotropic one, and $\alpha = \overline{n}/n_s = \overline{n}$ that is $n_s = 1$. It is seen from Fig. 3, that if the defect layer goes near the system center, the PBG is significantly widened. If the defect layer is nearby the CLC borders, the defect modes practically do not appear, and reflection is practically the same as for the case when there is no defect. The reflection at the defect mode increases for the light with the RCP if the defect shifts to the system center. However there is an asymmetry in respect to the distance from the CLC left or right border, i.e. the system possesses structural non-reciprocity and can be used as an all-optical diode if there is absorption (or gain). Thus, the defect location change in the system has essential influence on reflection, as well as on the other optical parameters at the defect mode. Some change of the defect mode wavelength takes place, too.

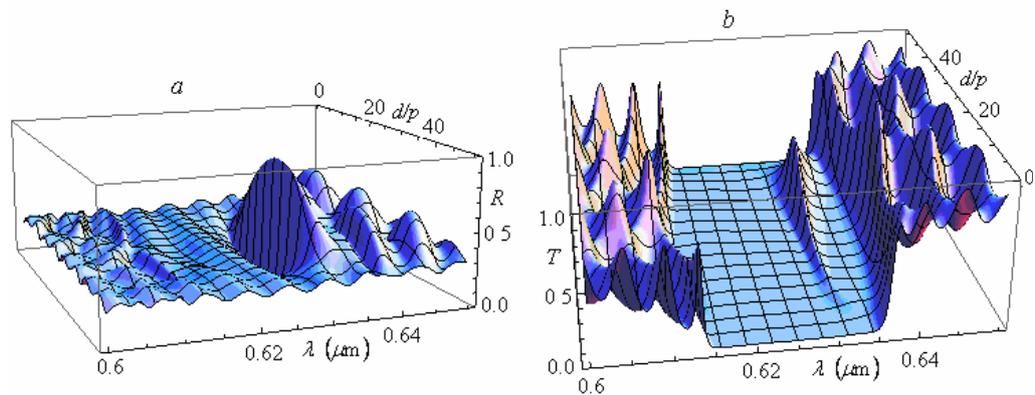

Fig. 3. The 3D plots of the reflection coefficient, $R$, (*a*) and the transmission coefficient, $T$, (*b*) on the wavelength, $\lambda$, and on the reduced distance between the CLC right border and the defect layer (on $d/p$, where $d$ is the distance between the CLC right border and the DL). The incident light is of the LCP (*a*) and RCP (*b*). The defect is isotropic with the thickness: $d^d = 0.1 \mu m$, and $\alpha = 1$. CLC layer thickness: *d=50p*.

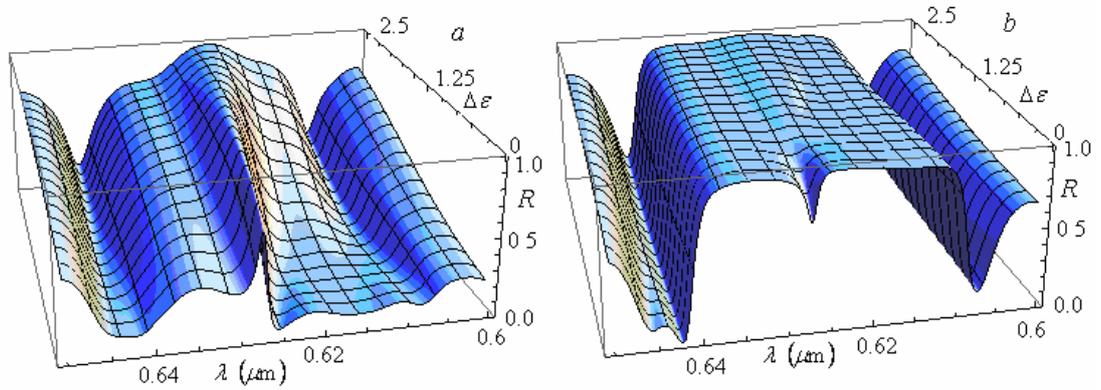

Fig. 4. The 3D plots of the reflection coefficient, *R*, on the wavelength, λ, and on the anisotropy . The incident light is of the LCP (*a*) and RCP (*b*). The defect thickness: $d^d = 0.1 \mu m$, and $\alpha = 1$. CLC layer thickness: *d=40p*.

As our calculations show, practically the same dependences are observed in the case of the structure with a thin anisotropic defect layer.

We pass on to investigation of the influence of the optical anisotropy of the defect layer on the reflection spectra. Representing the defect layer dielectric tensor principal values in the form, $\varepsilon_1^d = \varepsilon_0 + \Delta\varepsilon$ and $\varepsilon_2^d = \varepsilon_0 - \Delta\varepsilon$, we investigate the reflection spectra for various values of $\Delta\varepsilon$. In Fig. 4, the 3D dependences of the reflection coefficient, *R*, on the wavelength, λ, and on the anisotropy, $\Delta\varepsilon$ are represented. The incident light is LCP (a) and RCP (b). As it is seen from the figure, an increase in the anisotropy leads to an increase of reflection of the light in the PBG and to an increase of the defect mode width, for the light of the LCP, and, as our calculations show, further anisotropy increase leads to the 100% reflection in the PBG (in this case the defect layer becomes a half-wave plate). For the light of the RCP, the defect mode appears at smaller anisotropy values and the spectral hole diminishes steadily and eventually vanishes if the anisotropy increases (see, also [29]).

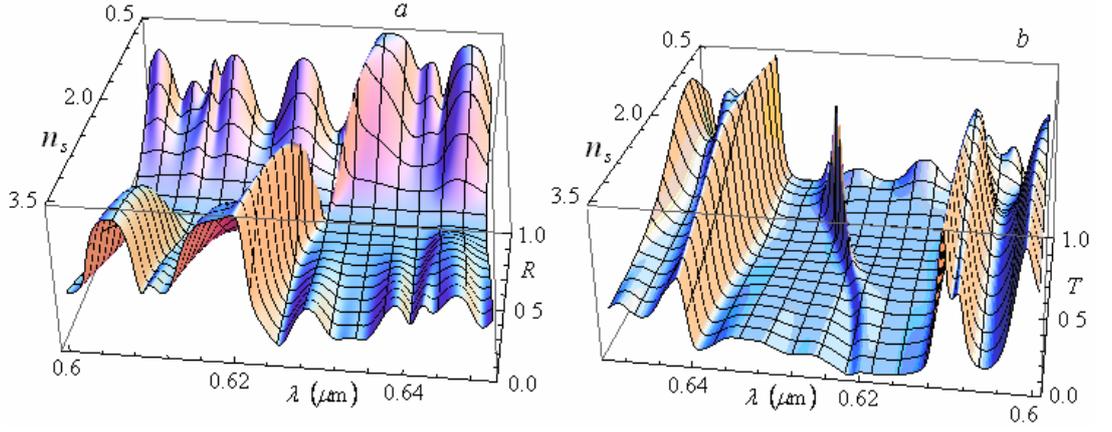

Fig. 5. The 3D plots of the reflection coefficient, *R*, (a) and the transmission coefficient, *T*, (b) on the wavelength, λ, and on $n_s$. The incident light is of the LCP (*a*) and RCP (*b*). The defect is isotropic with the thickness: $d^d = 0.1 \mu m$. CLC layer thickness: *d*=50*p*.

The change of the index, $n_s$, of the medium surrounding the system on both its sides has essential influence on the system reflection and the reflection at the defect modes. In Fig. 5 the dependences of the reflection coefficient, *R*, on the wavelength, λ, and on $n_s$ are presented. The incident light is LCP (a) and RCP (b). The defect is isotropic. As it is seen from the figure, the increase of the $n_s$ leads to strong reflection both inside and outside the PBG, as well as at the defect mode. This change is especially strong for LCP light. The increase in $|\bar{n} - n_s|$ also leads to a change in frequency of the defect mode in a significant interval. For an anisotropic defect and at $\bar{n} = n_s$, the defect mode for LCP light is absent, and it is natural, because a CLC with a defect in its structure is a micro-resonator, and one needs multi-reflections to arouse defect modes. As show in our calculations, practically the same dependences are observed in the case of anisotropic layer defect. In this case, the defect mode exists at $\bar{n} = n_s$, as well.

As it is mentioned earlier, the defect mode for the LCP incident wave reveals itself in the form of a peak in the reflection spectrum inside the PBG, but that for the RCP substantially reveals itself in the form of a hole in the reflection spectrum. Additionally, both polarizations have practically the same wavelength and the same reflection coefficient at the center of the peaks.

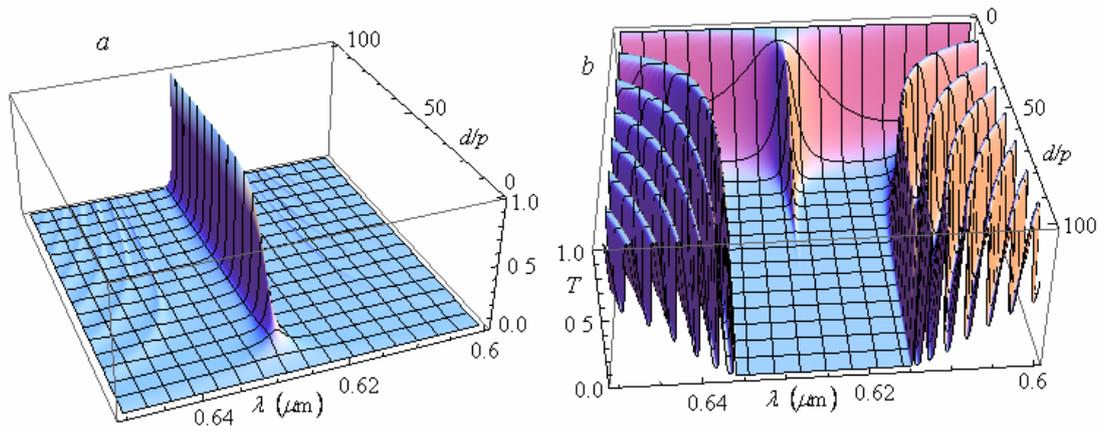

Fig. 6. The 3D plots of the reflection coefficient, *R*, (*a*) and the transmission coefficient, *T*, (*b*) on the wavelength, λ, and on the reduced thickness of the **CLC layer** (on *d/p,* where *d* is the **CLC layer** thickness). The incident light is of the LCP (*a*) and RCP (*b*). The defect is anisotropic with the thickness: $d^d = 0.1 \mu m$. $\alpha = 1$.

As shown in [22], for the case of a small CLC thickness a defect mode with the RCP is strongly excited, and it vanishes in the case of very large CLC thicknesses. The case for a defect mode with the LCP behaves the other way round; it is excited very little for small thicknesses, and appears very strongly at large thicknesses. In the case of the intermediate thicknesses they are excited almost equally. Let as note that an analogous behavior is observed for the CLC with a twist defect [30]. Here we present the results of a more detailed analysis of the behavior of the defect mode of various thicknesses in the CLC layer. In Fig. 6a we present plots of the reflectance coefficient, *R,* on the wavelength, λ, and on the reduced thickness of the CLC layer (on *d/p,* where *d* is the CLC layer thickness); and in Fig. 6b the transmittance coefficient, *T,* on λ and on *d/p* are presented. The incident light is of the LCP (a), and RCP (b). The defect is anisotropic. The same regularities are also observed for isotropic defects.

## IV. CONCLUSION

We showed that, in contrast to the anisotropic defect case, if the defect thickness changes, the changes of the defect mode half-widths become insignificant for the isotropic defect. This can enable strong light accumulation and low-threshold laser radiation at the defect mode for comparatively larger defect layer thicknesses, which can easier be carried out experimentally than for a thinner defect.

Indeed, as it is well known, the CLC doped with laser dyes (resonance molecules) can be used for designing feedback lasers without any mirror use. In an amplifying media (for instance, in CLCs doped with fluorescent guest-molecules, but in the way that the fluorescent peak is either in the PBG, or covers it), the PBG has significant influence on the radiation spectrum. The wave is evanescent (decreases exponentially) in the PBG and, consequently, the spontaneous radiation vanishes. The explanation is that the photonic density of states (PDS) vanishes and, as the spontaneous radiation intensity is proportional to PDS; the spontaneous radiation intensity also vanishes. At the PBG borders, the spontaneous radiation life time, $\tau_s$, sharply increases ($\tau_s$ decreases oscillating outside it) and makes the stimulated radiation strongly go up. Laser generation threshold energy essentially decreases, and the radiation essentially increases. And, as the CLC helix pitch can be changed, as well as tuned, a possibility of the laser radiation wavelength tuning arises, which can have most important practical significances. Yablonovitch has predicted that a low-threshold lasing will occur at defect modes within the band gap of PCs, too, since the excitation energy is not drained by spontaneous emission into the modes other than the lasing mode [31]. Lasing is further facilitated at the wavelength of the defect mode since the photon dwelling time is enhanced, giving ample opportunity for amplification by the stimulated emission.

We showed that the change of the refraction coefficient of the medium bordering the system on both sides leads to a significant change of the reflection at the defect mode, as well as to a defect mode wavelength change in a significant wavelength interval.

Also, let us note that the results obtained in this paper can be used for designing: narrow-band filters and mirrors, optical diodes, etc.